\documentclass[12pt]{article}

\usepackage{graphicx}
\usepackage{amsmath}
\usepackage{amssymb}

\usepackage{caption}
\usepackage{titlesec}
\titleformat{\section}
  {\normalfont\fontsize{15}{18}\bfseries}{\thesection}{1em}{}

\begin{document}

${}$\\
\begin{center}
\vspace{36pt}
{\large \bf Causal Dynamical Triangulations \\ \vspace{12pt} 
without Preferred Foliation}
\vspace{48pt}

{\sl S. Jordan}$\,^{a}$
and {\sl R. Loll}$\,^{a,b}$

\vspace{24pt}

{\footnotesize

$^a$~Radboud University Nijmegen, \\
Institute for Mathematics, Astrophysics and Particle Physics, \\
Heyendaalseweg 135, NL-6525 AJ Nijmegen, The Netherlands.\\
{email: s.jordan@science.ru.nl, r.loll@science.ru.nl}\\

\vspace{10pt}

$^b$~Perimeter Institute for Theoretical Physics,\\
31 Caroline St N, Waterloo, Ontario N2L 2Y5, Canada.\\
{email: rloll@perimeterinstitute.ca}\\

}

\vspace{48pt}

\end{center}

\begin{center}
{\bf Abstract}
\end{center}
We introduce a generalized version of the Causal Dynamical Triangulations (CDT) formulation of quantum gravity, 
in which the regularized, triangulated path integral histories maintain their causal properties, but do not have
a preferred proper-time foliation. An extensive numerical study of the associated nonperturbative path integral 
in 2+1 dimensions shows that it can nevertheless reproduce the emergence of an extended de Sitter universe 
on large scales, a key feature of CDT quantum gravity.
This suggests that the preferred foliation normally used in CDT is not a crucial (albeit convenient) part of its
background structure.

\newpage

\section{Quantum gravity and background independence}

In the context of nonperturbative quantum gravity, the mildest form of {\it background independence} one may impose
is that of the absence of a distinguished reference metric, like the Minkowski space of perturbation
theory. One may favour a more radical interpretation of background independence, removing further elements of
the classical formulation from the priors of the quantum theory: the differential structure of spacetime, its topology, or even
its dimension. Of course, there is a price to pay for such theoretical asceticism -- the less structure one assumes at the outset, the
harder one has to work to recover it dynamically from the quantum theory, as will be necessary if the quantum-gravitational
theory is to have a good classical limit.

No theory of quantum gravity can be completely background-free, since {\it some} choices must be made for
its degrees of freedom, symmetries and general principles such as causality or locality, to allow us to {\it calculate} 
anything. Different candidate theories differ in their choices \cite{qgcollection}. 
Rephrasing the issue of background independence, we may then ask 
how little background structure one can get away with in constructing a theory of quantum gravity, 
while maintaining a demonstrable link with classical general relativity. Of course, the influence of various 
aspects of background structure can be assessed meaningfully
only once a particu\-lar formulation is sufficiently complete to have produced numbers and results. 

This is arguably the case for ``Quantum Gravity from Causal Dynamical Triangulations (CDT)" \cite{physrep},
a nonperturbative implementation of the (formal) gravitational path integral
\begin{equation}
\label{eq:pigrav}
Z=\hspace{-.6cm}\mathop{\int}_{\mathrm{geometries\, [g]}}\hspace{-.7cm}{\mathcal{D}[g]}\exp(iS_{\mathrm{EH}}[g]),
\end{equation}
where $S_{\mathrm{EH}}[g]$ denotes the Einstein-Hilbert action as functional of the four-metric $g_{\mu\nu}$. CDT
uses a regularization of curved spacetime geometries $[g_{\mu\nu}]$ in terms of flat simplicial building blocks, such that 
eq.\ (\ref{eq:pigrav}) becomes a sum over triangulations. The notable results of this approach include 
the dynamical emergence (for suitable bare couplings, in the continuum limit) of a stable ground state of geometry, 
which on large scales is
four-dimensional \cite{cdt4d} and shaped like a de Sitter space \cite{desitter}, and on Planckian scales undergoes a 
dimensional reduction to a value compatible with 2 \cite{spectral}. The underlying statistical model has recently been
shown to possess a second-order phase transition, so far unique in nonperturbative quantum gravity \cite{trans}.  

\begin{figure}[t]
\centerline{\scalebox{0.43}{\includegraphics{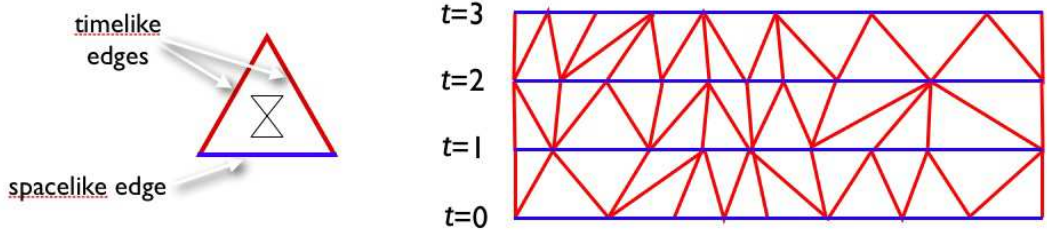}}}
\caption{CDT in 1+1 dimensions has a single type of building block, a flat Minkowskian triangle (left, with
light cone indicated). The triangles are glued in a sequence of strips, giving
a piecewise flat Lorentzian manifold (right). Each one-dimensional slice at integer $t$ consists of spacelike
edges only.}
\label{fig:2dCDTstrip}
\end{figure}

CDT is background-independent in the weak sense: its regularized path integral is a ``democratic" sum
over Lorentzian spacetimes, without singling out any particular one. Its remaining background structure is fairly 
minimalist: the path integral depends on the local Einstein-Hilbert Lagrangian and has no gauge redundancies,
because it is defined in purely geometric terms without ever introducing coordinates. The triangulated
spacetimes to be summed over are four-dimensional simplicial manifolds, which share a fixed topology 
of product type $I\!\times\! {}^{(3)}\Sigma$, for some compact $\! {}^{(3)}\Sigma$, usually taken to be $S^3$.
Importantly, they also share a discrete notion of a ``proper time" $t$, which is related to the fact that
they have a product structure {\it as triangulations}. This means that each triangulation
can be viewed as a sequence $t=0,1,2,..., T$ 
of three-dimensional {\it spatial} triangulations of topology $\! {}^{(3)}\Sigma$, where
the ``space" between each pair of consecutive slices is ``filled in" by a layer 
$\Delta t\! =\! 1$ of four-dimensional Lorentzian simplices, such that the stack of $T$ layers forms a simplicial
manifold \cite{physrep}. Fig.\ \ref{fig:2dCDTstrip} illustrates the analogous situation in 1+1 dimensions.

The question then arises which -- if any -- of these background structures may be weakened or dropped, while 
keeping the results mentioned above, certainly those pertaining to the classical limit. We already know
that summing over Euclidean {\it spaces} instead of Lorentzian {\it spacetimes} of fixed topology in the
regularized sum over triangulations does {\it not} yield an extended ground state 
geometry \cite{Ambjorn:1991pq,Agishtein:1992xx,Catterall:1994pg}; this is what motivated using Causal
Dynamical Triangulations instead of (Euclidean) Dynamical Triangulations (DT) in the first place. 

The {\it causal} nature of CDT refers to the fact that individual
path integral configurations have a well-defined local light cone structure and that there are no closed
timelike curves. In practice, the latter requirement is enforced by organizing the four-dimensional simplices
into the product structure described above. In this way the notion of causality -- at the kinematical level --
becomes tied to a
specific substructure of the triangulated manifolds, with an associated preferred notion of (discrete) time $t$.
In the continuum limit, there
is strong evidence from the matching with de Sitter space \cite{desitter} that $t$ can be identified with
continuum proper time in an averaged sense {\it on large scales}, but there is no a priori claim that $t$
is associated with a physically meaningful notion of continuum time {\it locally}. 
 
Until now it has not been clear how to dissociate the causal structure of CDT from the preferred discrete
foliation of the individual path integral histories, while staying in the purely geometric set-up of dynamical
triangulations. We explain in Sec. 2 how to modify the CDT framework to achieve this. Previous attempts
have been made to ``relax" the proper-time slicing in CDT: 
in \cite{markopoulousmolin} an additional, variable ``lapse label" was associated with 
given, foliated CDT configurations in 1+1 dimensions, and interpreted as a local rescaling of time.
Imposing some bounds on the average lapse, it was argued that this does not affect the 
continuum limit. Reference \cite{konopka} considered similar ideas for CDT in 2+1 dimensions, and
suggested the use of generalized building blocks which extend over more than one layer, for example,
over $\Delta t\! =\! 2$.

Below, we will report on the results of a full-fledged numerical simulation of
a generalized version of CDT quantum gravity in 2+1 dimensions, whose configurations do not have a preferred 
foliation in time. 
Nevertheless, as we shall see, there is a whole region in the phase space of the generalized model where 
large-scale results compatible with standard CDT are reproduced. This provides definite evidence that 
the presence of a preferred time slicing in Causal Dynamical Triangulations is not an essential element of its 
background structure.

\begin{figure}[t]
\centerline{\scalebox{0.43}{\includegraphics{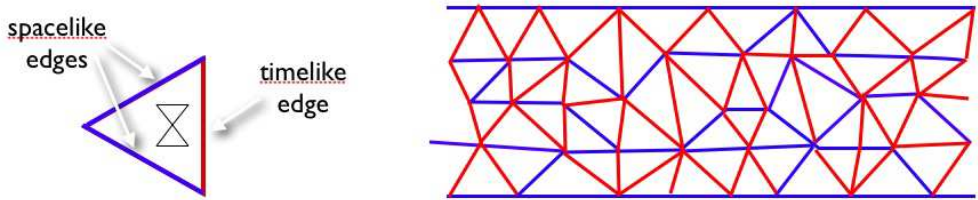}}}
\caption{The addition of a new Minkowskian building block in CDT in 1+1 dimensions (left) allows for the construction of
piecewise flat Lorentzian manifolds without a preferred foliation in time (right).}
\label{fig:2dgCDTstrip}
\end{figure}

\section{CDT without foliation: theoretical aspects}

The key idea in generalizing the usual CDT setting is to stick to the usual two (squared) 
edge lengths, $\ell^2_s\! =\! a^2$ for all 
spacelike links and $\ell^2_t\! =\! -\alpha a^2$ for all timelike links, $\alpha\! >\! 0$, but to add new Minkowskian building blocks 
compatible with these assignments. For illustration, Fig.\ \ref{fig:2dgCDTstrip} shows the situation in dimension
1+1, where one can have one additional type of flat Lorentzian triangle, with one time- and two spacelike edges.

\begin{figure}[t]
\centerline{\scalebox{0.7}{\includegraphics{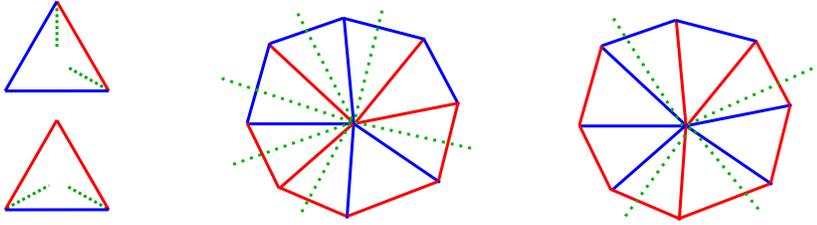}}}
\caption{The two Minkowskian triangles allowed in 1+1 dimensions; dotted lines indicate light rays through the corner
points (left). Gluing these together can result in local causality violations, for example, too many light cones meeting at
a vertex (centre). At a causally well-behaved vertex, one crosses exactly four lightlike direction when going
around the vertex (right).    
}
\label{fig:causality2d}
\end{figure}

Arbitrary gluings of the enlarged set of triangles will in general violate causality already locally, in the sense that 
the light cone structure at a vertex may be pathological. To avoid these situations, we require that when going around
a vertex once, lightlike directions are crossed exactly four times (Fig.\ \ref{fig:causality2d}). 
One can show that on a contractible piece of CDT triangulation with two spatial boundaries in 1+1 dimensions 
this condition is sufficient to prohibit closed timelike curves.\footnote{We thank R.\ Hoekzema and S.\ Smith for
discussions on this point.}
We will in the following consider only configurations which are causal in this strong,
global sense. In the concrete case of generalized CDT in 2+1 dimensions with spherical slices, the absence of closed
timelike curves seems to be implied by requiring local ``link causality", a higher-dimensional version of the vertex causality
depicted in Fig.\ \ref{fig:causality2d}, although we do not have an analytic proof at this stage (see \cite{forthcoming} for
more details). As illustrated by
Fig.\ \ref{fig:2dgCDTstrip}, general causal gluings of the two types of triangles do {\it not} possess the distinguished
proper-time slicing present in standard CDT, although one can clearly identify many piecewise straight
spatial slices in them, for example, by following suitable chains of spacelike links. 

Analogous kinematical set-ups exist in any dimension and can be used to define generalized ensembles on which to
evaluate the nonperturbative quantum-gravitational path integral. We will focus here on the case of 2+1 dimensions,
because in standard CDT its large-scale features are closely analogous to those of the physical theory 
(one finds an emergent, {\it three}-dimensional de Sitter universe \cite{3dcdt,benedettihenson}), and
because the complexity of the 
numerical implementation of the path integral, although considerable, is less formidable than in 3+1 dimensions. 
For the purpose of this letter, we confine ourselves to a description of the main results; many more details on
the theoretical and numerical aspects of the model can be found in \cite{forthcoming}.

\begin{figure}[t]
\centerline{\scalebox{0.55}{\includegraphics{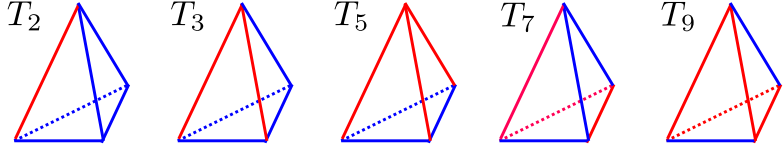}}}
\caption{All three-dimensional Minkowskian tetrahedra that can be constructed using only two types of edges,
spacelike (blue) and timelike (red). The $T_7$-tetra\-hedron is only Lorentzian for $\alpha\! <\! 1$.}
\label{fig:tetrahedra}
\end{figure}

In 2+1 dimensions, eleven types of tetrahedra can be built from the two fixed edge lengths $\ell_s$ and $\ell_t$ 
introduced earlier. Four
of them (denoted by $T_2$, $T_3$, $T_5$ and $T_9$) have the correct Minkowskian signature for all values of the
asymmetry parameter $\alpha$, a fifth one ($T_7$) only for $\alpha\! <\! 1$, see Fig.\ \ref{fig:tetrahedra}.
For simplicity, we will study the model based on the first four tetrahedra. Note that standard CDT uses only tetrahedra 
of types $T_5$ and $T_9$. Like in standard CDT, performing the Wick rotation to convert the regularized path integral 
to a real partition function (and make it amenable to Monte Carlo simulations) amounts to an analytic continuation of
the asymmetry parameter $\alpha $ to $-\alpha$. The presence of the new building blocks leads to an upper bound
on $\alpha$ for the Wick rotation to exist, in addition to the usual lower bound, yielding $1/2\! <\alpha\! < 3$.
After the Wick rotation to Euclidean signature, the discretized Einstein-Hilbert action can be written in a form 
particularly useful in simulations, namely,
\begin{displaymath}
\label{eq:linearaction2}
S^\mathrm{eucl}=\widetilde{c_1} N_0 + \widetilde{c_2} N_3 + \widetilde{c_3} N_3^{T_2} + \widetilde{c_4} 
N_3^{T_3} + \widetilde{c_5} N_3^{T_5},\quad 
\widetilde{c_i}=\widetilde{c_i}(k,\lambda,\alpha),
\end{displaymath}
where
$N_0$ and $N_3$ are the total numbers of vertices and tetrahedra in a given triangulation, $N_3^{T_i}$ is the total number 
of tetrahedra of type $T_i$, and $\lambda$ and $k$ are the bare cosmological and bare inverse gravitational couplings. 
The explicit expressions for the $\widetilde{c_i}(k,\lambda,\alpha)$ 
are given in \cite{forthcoming}, and are not important here.

\section{CDT without foliation: numerical set-up}

To study the nonperturbative properties of the new, generalized CDT path integral, we have defined a set of Monte Carlo
moves for use in the numerical simulations. We work with compact spatial slices of the topology of a two-sphere. 
Since the usual choice of compactifying the time direction to a circle led to severe thermalization problems, we work with
the overall topology of a three-sphere, whose south and north pole can be thought of as degenerate spacelike boundaries
at the beginning and end of time.
Since the spatial volume has a tendency to become very small for early and late times, this choice will not significantly
influence our results.

Next, we have to face the fact that getting rid of a preferred notion of time we also lose direct access to the average
``volume profile" (two-volume $N_2$ as a function of time), the large-scale observable instrumental in connecting CDT to a de Sitter 
minisuperspace cosmology \cite{3dcdt,benedettihenson}. To reintroduce a notion of time on our universes we proceed
as follows.

Given a vertex $v$, consider the set of all future-oriented paths connecting $v$ with the 
north pole. The number of links in each path defines a distance between $v$ and the north pole. By averaging
this quantity over all paths we obtain an average distance $d_f$. Repeating the procedure for past-oriented paths, connecting
$v$ to the south pole, gives another average distance $d_p$. The time coordinate of $v$ is then defined as 
$t\! =\! d_f-d_p$. Note that for foliated CDT configurations, this coincides with the usual discrete proper time, up to a trivial factor.

Next, define a {\it spatial slice} as any subset of spatial triangles forming a two-sphere, such that by cutting along the 
sphere the spacetime triangulation decomposes into two disconnected parts.
The time coordinate we assign to such a slice is the average of all time coordinates of its vertices. --
We now we have all ingredients to measure the desired volume profiles. 
Since the number of spatial slices of an individual path integral configuration can become very large, we use a 
statistical method to generate a subset of spatial slices which is evenly distributed along the time direction. 
In order to perform an ensemble average of the volume distribution we use a nontrivial averaging 
algorithm \cite{forthcoming}.

\section{CDT without foliation: results on semiclassicality}

\begin{figure}[t]
\centerline{\scalebox{0.75}{\includegraphics{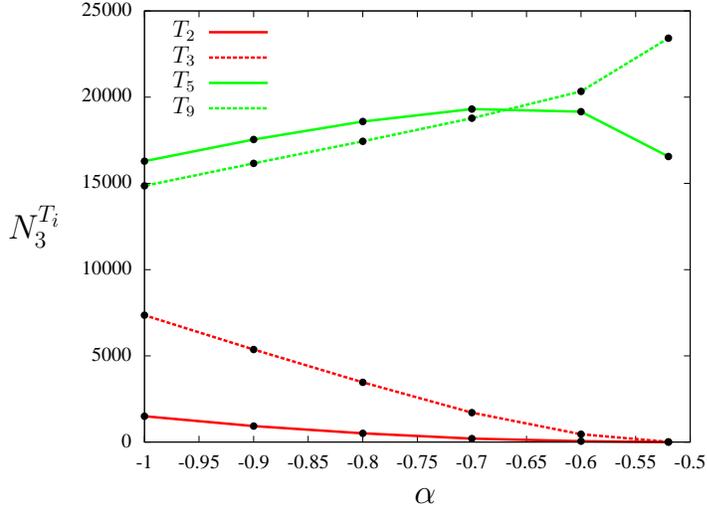}}}
\caption{The numbers $N_3^{T_i}$ of tetrahedra of type $i$, averaged over the sampled triangulations, as function of the 
coupling $\alpha$, for $k\! =\! 0$ and at system size $N_3\! =\! 40$k. The green curves represent the tetrahedral types used in 
standard CDT.}
\label{fig:n3_alpha}
\end{figure}

Since we work at constant total volume $N_3$, as usual in CDT, the phase diagram of our model is spanned by the 
parameters $k$ and $\alpha$. Using our newly written simulation software, we have investigated a range of
$k$-values (for fixed $\alpha\! =\! -1$) and the 
interval $-1\! <\! \alpha\! <\! -1/2$ for fixed $k\! =\! 0$. Note that $\alpha\! =\! -1/2$ defines a boundary of the phase diagram, 
due to the Wick-rotation condition mentioned earlier. 

We have counted the tetrahedra of the different types at various points, to identify regions where there are
very few building blocks of the new types $T_2$ and $T_3$. This does not seem to happen for points
with fixed $\alpha\! =\! -1$ and $-1\leq \! k\!\leq\! 0$, but the situation is different along the $\alpha$-axis 
(i.e. for $k\! =\! 0$), see 
Fig.\ \ref{fig:n3_alpha}. The numbers $N_3^{T_2}$ and $N_3^{T_3}$ go to zero as we approach the phase 
boundary at $\alpha\! =\! -1/2$, which means that near the boundary the triangulations are 
almost perfectly foliated. At the effective level the dynamics 
should therefore be close to that of 2+1 dimensional foliated triangulations in the extended 
phase \cite{3dcdt}, an expectation we will confirm in what follows.

\begin{figure}[t]
\centerline{\scalebox{1.0}{\includegraphics{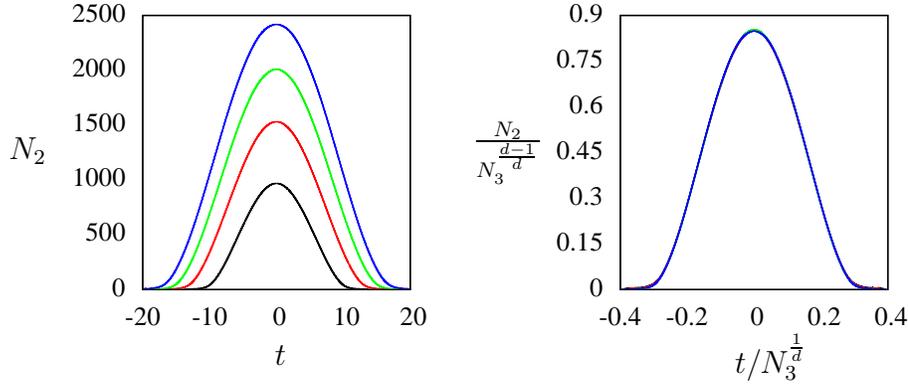}}}
\caption{The average volume profiles $\langle N_2(t)\rangle$ at $(k,\alpha)\! =\! (0.0,-1.0)$ for system sizes $N_3\! =\! 40, 
80, 120$ and $160\mathrm{k}$. Data shown on the left are unscaled, while those on the right have been rescaled
with appropriate powers of $N_3$, using $d\! =\! 2.91$ to achieve a best overlap.}
\label{fig:vpfss_k0.0}
\end{figure}

To study the volume profiles, we have performed simulations at system sizes $N_3\! =\! 40$, 80, 120 
and 160k, at various points in phase space. Fig.\ \ref{fig:vpfss_k0.0} (left) shows the average volume 
profiles for these system sizes at $(k, \alpha)\! =\! (0,-1)$, exhibiting similar-looking, localized distributions of 
volume. This clearly 
shows that the geometries are macroscopically extended. Similar results were found at all points with $k\! =\! 0$ we 
have investigated. For fixed $\alpha\! =\! -1$ we observe the presence of extended geometries in the region 
$k\!\lesssim\! 0.25$. 
At $k\!\approx\! 0.25$ there is a phase transition to a phase in which the geometry becomes a long, thin tube in the time 
direction, with minimal spatial volume almost everywhere.

We have performed a finite-size scaling analysis of the average volume profiles, analogous to what was done in 
\cite{desitter} for foliated triangulations in 3+1 dimensions. Assuming the average geometry has a macroscopic dimension
$d$, we expect time intervals to scale like $N_3^{1/d}$ and spatial volumes like $N_3^{(d-1)/d}$. 
When plotting the profiles with axes rescaled accordingly, they should fall on top of each other. 
Using an algorithm which estimates $d$ from finding the best overlap of the average volume 
profiles \cite{forthcoming}, we have found $d\! =\! 2.91$ at $(k, \alpha)\! =\! (0,-1)$. Fig.\ \ref{fig:vpfss_k0.0} 
shows that for this value of $d$ the overlap is almost perfect.

\begin{figure}[t]
\centerline{\scalebox{1.1}{\includegraphics{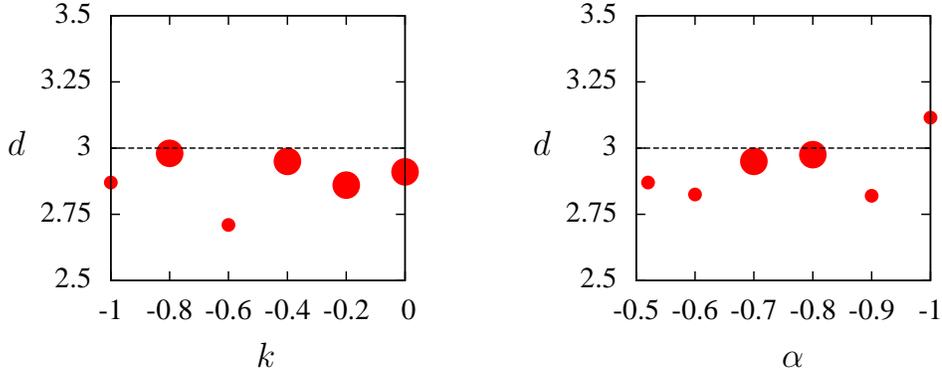}}}
\caption{Estimates for the macroscopic dimension $d$ from finite-size scaling, from measurements performed for fixed 
$\alpha\! =\! -1.0$ (left) and fixed $k\! =\! 0.0$ (right). Large dots represent measurements where the volume 
profiles overlap with excellent quality, smaller dots indicate lesser quality.}
\label{fig:d}
\end{figure}

We have repeated the analysis for the other selected points in phase space. Fig.\ \ref{fig:d} shows the calculated 
estimates for the macroscopic dimension $d$, from measurements taken for fixed $\alpha$ and variable $k$ and vice versa.
The size of the dots on the figures reflects a qualitative assessment of how well the volume profiles overlap at that particular
phase space point, where larger dots indicate excellent quality (and therefore higher reliability of the data point), while
smaller dots stand for overlaps of a somewhat lesser quality.\footnote{We have not included error bars, because of 
systematic errors we currently cannot estimate quantitatively, see \cite{forthcoming} for further discussion.} 
We observe that the six high-quality measurements 
yield macroscopic dimensions between $d\! =\! 2.85$ and $d\! =\! 3.00$, while the remaining data points display a wider 
spread. This represents good evidence that the extended universe we observe in the simulations has 
macroscopic dimension three.

\begin{figure}[t]
\centerline{\scalebox{0.75}{\includegraphics{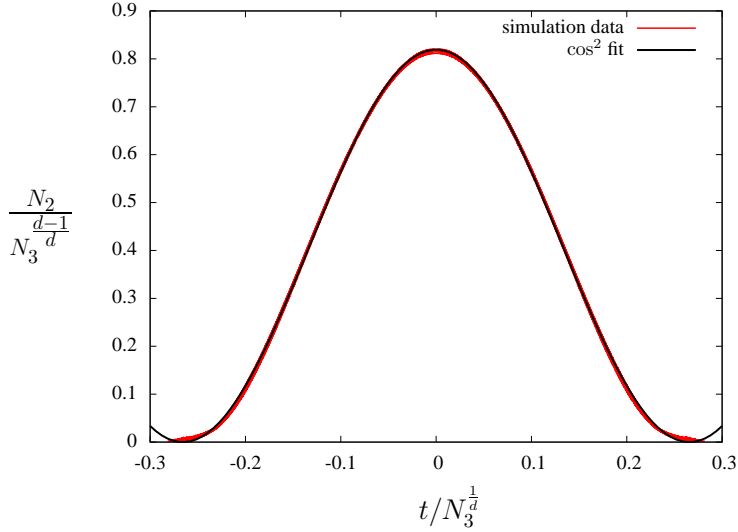}}}
\caption{Rescaled average volume profile at $(k,\alpha)\! =\! (-0.8,-1.0)$ and best fit using the $\cos^2$-ansatz, 
corresponding to $d\! =\! 2.98$.}
\label{fig:fit_k-0.8}
\end{figure}

To further strengthen the case for semiclassicality of the emergent spacetime on large scales,
we have matched the shape of the average volume profile to that of a classical Euclidean de Sitter space, 
in other words, a round three-sphere, like what has been done successfully for foliated CDT in 2+1 and 3+1 
dimensions. The volume profile as a function of proper time has the functional form $V_2(t)=a \cos^2(b t)$, for 
constants $a$ and $b$. Selecting those phase space points at which the overlaps were
excellent, we then attempted to fit the corresponding curves using the $\cos^2$-ansatz with two free parameters.
For illustration, Fig.\ \ref{fig:fit_k-0.8} shows the result at the point $(k,\alpha)\! =\! (-0.8,-1.0)$. Obviously, the 
only relevant part of the fit function is the region between the two minima. The $\cos^2$-ansatz fits the average volume 
profile almost perfectly, including the distance in time between the two minima, with a similar situation found at other phase 
space points. This reconfirms the three-dimensionality of the extended universe, and provides very good evidence
that it is of de Sitter type on large scales, just like in the standard CDT formulation  of quantum gravity.  

\section{Discussion}

In this letter we have shown how to generalize the standard formulation of Causal Dynamical Triangulations, 
such that at the level of the regularized theory its causal properties are left intact, but without tying them
to any preferred lattice substructure in the form of a foliation. Key to implementing this program was the
introduction of additional Minkowskian building blocks (combined with a local causality constraint), 
such that generic causal triangulations no longer 
have a foliated structure. Relaxing the foliation of CDT could a priori have led to an instability of the 
system, leading to the disappearance of the semiclassical behaviour seen in the extended phases of
higher-dimensional CDT. Our nonperturbative simulations of the ensemble of nonfoliated CDT geometries 
have shown that this does not happen, at least not in 2+1 dimensions. On the contrary, we have found 
an extended region in the phase space of the generalized model where the geometries preferentially 
arrange themselves to be ``almost foliated". By explicit construction, we showed that by reintroducing notions 
of ``time" and ``spatial
slices" through appropriate averaging, the semiclassical results of standard CDT can be reproduced in
this region of phase space, not just qualitatively, but quantitatively. 

Conjecturing that our findings will carry over to 3+1 dimensions and are an indication that the two models --
with and without preferred foliation -- describe the same continuum physics, 
we conclude that the discrete proper-time foliation of CDT is dispensable as part of its ``background structure".
This clarifies further the role of the discrete 
label ``$t$" in causal triangulations, and strengthens earlier findings \cite{markopoulousmolin,konopka}, this time in
the context of a fully worked-out, dynamical analysis of a definite DT model. 
On the other hand, we have got a glimpse of the considerable technical complications that arise
when abandoning the foliation. If our conjecture is correct, taking this step is not actually necessary, which
may make sticking with the original CDT framework preferable from a practical point of view.

\vspace{.5cm}

\noindent {\bf Acknowledgements.} 
The authors' contributions are part of the research programme of the Foundation for Fundamental Research 
on Matter (FOM), financially supported by the Netherlands Organisation for Scientific Research (NWO).
The work was also sponsored by NWO Exacte Wetenschappen (Physical Sciences) for the use 
of supercomputer facilities, with financial support from NWO. Research at Perimeter Institute is supported by the 
Government of Canada through Industry Canada and by the Province of Ontario through the Ministry of Economic 
Development and Innovation.

\end{document}